\documentclass[10pt, twocolumn]{IEEEtran}
\usepackage{amsfonts, amsmath, amssymb, epsfig, subfigure, graphicx}

\newtheorem{theorem}{Theorem}

\newtheorem{corollary}[theorem]{Corollary}

\newcommand{\thmend}{\hspace*{\fill}~\QEDopen\par\endtrivlist\unskip}
\newcommand{\eqdef}{ \stackrel{\bigtriangleup}{=} }

\begin{document}
\title{The Multiplexing Gain of MIMO $X$-Channels with Partial Transmit Side-Information}
\author{Natasha Devroye and Masoud Sharif\\
{\it Harvard University, Boston University}}
%\thanks{Natasha Devroye is with the Division of Engineering and Applied Sciences, Harvard
%University (e-mail: ndevroye@deas.harvard.edu). Masoud Sharif is with the Department of Electrical and Computer Engineering, Boston University (e-mail:sharif@bu.edu). }
\maketitle
\begin{abstract}
THIS PAPER IS ELIGIBLE FOR THE STUDENT PAPER AWARD.
In this paper, we obtain the scaling laws of the sum-rate capacity of a MIMO X-channel, a 2 independent  sender, 2 independent receiver channel with messages from each transmitter to each receiver, at high signal to noise ratios (SNR). The X-channel has sparked recent interest in the context of cooperative networks and it encompasses the interference, multiple access, and broadcast channels as special cases. Here, we consider the case with partially cooperative transmitters in which \emph{only} partial and asymmetric side-information is available at one of the transmitters. It is proved that when there are $M$ antennas at all four nodes, the sum-rate scales like $2M\log {\rm SNR}$ which is in sharp contrast to $\left[\lfloor \frac{4M}{3} \rfloor,\frac{4M}{3}\right] \log{\rm SNR}$ for non-cooperative X-channels \cite{maddah-ali,jafar_degrees}. This further proves that, in terms of sum-rate scaling at high SNR, partial side-information at one of the transmitters and full side-information at both transmitters are equivalent in the MIMO $X$-channel. 
\end{abstract}
\section{Introduction}
Cooperation in wireless networks has sparked much recent interest in the research community.  Having wireless nodes cooperate at the transmitting and/or receiving end, can, for example, improve rates, diversity, or  power utilization.  How much gain cooperation provides depends on a number of factors, including the signal to noise ratio (SNR) and the amount of side information at the transmitters. In this work, we look at the value of partial transmitter cooperation, in terms of the scaling law of the sum-rate in the MIMO $X$-channel at high SNR .\footnote{Although the achievable rate region we derive is for general channels, the multiplexing gain results hold only for Gaussian noise channels, hence we will often omit the term Gaussian for brevity.}

%The multiplexing gain of a MIMO $X$-channel with \emph{no} transmitter cooperation  has been shown to lie in the range $\left[\lfloor \frac{4M}{3} \rfloor,\frac{4M}{3}\right]$ as the SNR $\rightarrow \infty$ \cite{maddah-ali, jafar_degrees}. 

The multiplexing gain of the sum-rate of the MIMO $X$-channel has been recently studied \cite{maddah-ali, jafar_degrees, jafar_interference}. The MIMO $X$-channel is a simple 2 transmitter (2 Tx), 2 receiver (2 Rx) channel in which each Tx has a message for each Rx. Its study is of theoretical interest, as it encompasses classical channels such as the interference, the multiple access and the broadcast channels. While the classical MIMO $X$ channel forbids  cooperation between the two Tx, in this work we allow for transmitter cooperation, in the form of transmitter \emph{side-information.}    

\begin{figure}
\centerline{\epsfig{figure=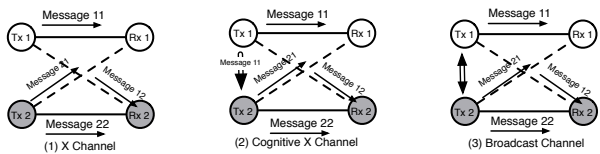, width=8cm}}
\caption{Three $2\times 2$ MIMO $X$-channels whose multiplexing gains are contrasted here. All Rxs decode independently.  (1) MIMO $X$-channel with one message from each Tx to each Rx and no side-information. (2) MIMO cognitive $X$-channel with partial asymmetric side-information:  Tx 2 knows message 11 but not message 12, and Tx 1 knows none of Tx 2's messages. (3) MIMO Broadcast channel: $2M$ transmitting antennas delivering messages to two independent  $M$ antenna Rxs.}
\label{fig:3Xchannels}
\end{figure}

The three channels shown in Fig. \ref{fig:3Xchannels} illustrate three multiuser MIMO channels with different amounts of transmitter side-information. In Fig. \ref{fig:3Xchannels}(1), the MIMO $X$-channel is illustrated\footnote{Throughout this work the term MIMO will denote that  there are $M$ antennas at each of the four nodes Tx 1, Tx 2, Rx 1 and Rx 2. In order to simplify results we will sometimes set $M=1$, in which case we drop the word MIMO.}.  The two transmitters, Tx 1 and Tx 2 operate independently, and each wishes to send messages to each of the two non-cooperating receivers Rx 1 and Rx 2. The messages are numbered 11, 12, 21 and 22, with the convention that indices will follow the form (Tx\# Rx\#).  We note that the MIMO interference channel is embedded in the MIMO $X$ channel and may be obtained by eliminating the cross-over messages 12 and 21.  The messages 11, 12, 21 and 22 will be simultaneously transmitted at the rates $R_{11}, R_{12}, R_{21}$ and $R_{22}$ 
respectively. The sum-rate achieved by this tuple is defined to be $R\eqdef R_{11}+R_{12}+R_{21}+R_{22}$. 

In Fig. \ref{fig:3Xchannels}(2), the MIMO cognitive $X$-channel is depicted. We define this to be a channel in which Tx 2 knows message 11, but not message 12, of Tx 1. Tx 1 knows none of Tx 2's messages. We use the term \emph{partial side-information} to refer to the the fact that Tx 2 knows only one of Tx 1's messages, and the term \emph{asymmetric side-information} to refer to the fact that Tx 2 has side-information regarding Tx 1 but not vice versa.  We note that the side-information could have been both message 11 and 12 or just message 12 at Tx 2.  In the latter event, analogous results are obtained by permuting the indices in our results. In the former, the rates achieved could be improved at low-medium SNR, but as we will see, the extra information contained in knowing the message 12 as well as message 11 is unnecessary as SNR $\rightarrow \infty$, where we will show that the partial information is sum-rate scaling optimal. In Fig. \ref{fig:3Xchannels}(3) shows a MIMO broadcast channel in which both Txs share all their messages (full, symmetric side-information) \cite{weingarten, shamai_achievable} with $2M$ transmit antennas and two independent Rxs with $M$ antennas each.

The multiplexing gain of the channel depicted in Fig. \ref{fig:3Xchannels}$(1)$ is defined as the limit of the ratio of the maximal achieved sum-rate $R_i$ in the capacity region ${\cal C}_i$, to the $\log$(SNR)\footnote{Note that the usual factor $\frac{1}{2}$ is omitted in any rate expressions, but rather the number of times the sum-rate looks like $\log(SNR)$ is the multiplexing gain.} as the SNR $\rightarrow \infty$, or
 \[M_i = \lim_{\mbox{{\small SNR}}\rightarrow \infty} \frac{\max R_i(\mbox{SNR})}{\log(\mbox{SNR})}.\]
%In this paper, we will examine the multiplexing gains achieved in these six channels. Specifically,  past research has shown the intuitive fact that the capacity regions of these six channels, denoted by ${\cal C}_1, {\cal C}_2, \cdots {\cal C}_6$ follow the rules ${\cal C}_1 \subset {\cal C}_2 \subset {\cal C}_3$ and ${\cal C}_4 \subset {\cal C}_5 \subset {\cal C}_6$.  The same is naturally true of the multiplexing gains of the six regions: $M_1 \leq M_2 \leq M_3$ and $M_4\leq M_5\leq M_6$, where
%The multiplexing gain of a channel corresponds, roughly and intuitively, to the number of parallel streams of information  one may transmit over that channel. At high SNR, noise is no longer the limiting factor. Rather, when sending multiple streams, it is the unmitigated interference between these streams that will limit the sum-rate capacity as the SNR $\rightarrow \infty$. 
The multiplexing gain of the channel of Fig. \ref{fig:3Xchannels}(1) has been recently studied \cite{jafar_degrees, maddah-ali}. The MIMO interference channel embedded in channel (1) is known to achieve  a multiplexing gain of $M$ \cite{jafar_interference}. This channel has no cross-over messages (messages 12 and 21). Interestingly, when cross-over information is present, as in the MIMO $X$-channel of (1),  the multiplexing gain lies in the range $[\lfloor \frac{4}{3}M\rfloor, \frac{4M}{3}]$ \cite{jafar_degrees, maddah-ali}, possibly improving upon the MIMO interference channel.  The MIMO broadcast channel of Fig. \ref{fig:3Xchannels}(3) is known to have a multiplexing gain of $2M$, equal to the number of transmit, and receive antennas. In the broadcast channel, full, symmetric Tx side-information is used, that is, both Tx 1 and Tx 2 know \emph{all} messages of the other. In this work, we show that such full transmit side-information is not strictly necessary to achieve the same sum-rate scaling of $2M$. We show that the partial side-information, shown in the MIMO cognitive $X$-channel also achieves a sum-rate scaling of $2M$ as SNR $\rightarrow \infty$.
%This shows that although partial side-information may increase achievable rates in  the interference channel in a medium ${\rm SNR}$-regime
%\cite{devroye_IEEE, jovicic}, at high ${\rm SNR}$, one cannot
%improve the scaling law of the sum-rate.
%In contrast, we will show that the cognitive $X$ channel achieves the same multiplexing gain as the broadcast channel. This is remarkable, as only partial asymmetric transmitter cooperation is assumed, in contrast to the full transmitter cooperation employed in the broadcast channel. 
%We will use the term \emph{cognitive} and \emph{asymmetric side-information} interchangeably. They denote a channel in which one transmitter's message is known by the other transmitter, but not vice versa. 
%The broadcast channel as shown in Fig. \ref{fig:6channels} (3)), as well as  the broadcast $X$ channel (Fig. \ref{fig:6channels}(6)) achieve multiplexing gains of $2$ for the single antenna case (notice that here the transmitter actually has two antennas that may fully cooperate, while each of the two receivers has a single antenna) \cite{jindal_highSNR}. In the case of $M$ antennas at all nodes, the multiplexing gain is $2M$, the same as if all receivers could cooperate. %Similarly, the multiplexing gain of a MIMO Multiple Access Channel (MAC) with M antennas at the two non-cooperating transmitting nodes and $M$ antennas at the single receiver, is known to be $M$. 

In this work, we assume \emph{non-causal} side-information, which may be either symmetric or asymmetric, but requires that all messages are known fully before transmission starts. This is in sharp contrast to the work \cite{host_madsen_coop}, in which similar 2 Tx, 2 Rx channels with causal side-information are considered. 
%There, side-information is passed through noisy channels, and the multiplexing gain is shown to be maximally 1 when $M=1$. 
%The multiplexing gain of 2 input, 2 output channels like the 6 ofFig. \ref{fig:6channels} has been considered in cooperative scenarios by the works \cite{host_madsen_int, host_madsen_coop}. There, the multiplexing gain of such channels is studied under the assumption that some of the transmitters or receivers may cooperate through noisy links, as opposed to our idealized, non-causal links (they may be thought of as having infinite capacity). Specifically, in \cite{host_madsen_int}, it is shown that a 2 sender, 2 receiver interference channel cannot have a multiplexing gain of more than 1, even if there are noisy links between the transmitters or receivers. 
In \cite{host_madsen_coop} it is shown that even if both the Txs and Rxs may cooperate using noisy links, in a causal fashion (although full duplex transmission is permitted) the multiplexing gain is always limited to 1 when $M=1$. This is in sharp contrast to when non-causal side-information is present, resulting in a multiplexing gain of 2.
% or perfect cooperation scenario, where joint encoding or joint decoding is made possible by perfect cooperation, or infinite links between either the senders or the receivers. 

This paper is structured as follows. 
%In section \ref{sec:cog_int} we briefly define the cognitive interference channel and then demonstrate that in the single antenna case, the multiplexing gain of the sum-rate is 1. 
In section \ref{sec:cog_X} we define the MIMO cognitive $X$ channel and demonstrate an achievable rate region, which we to show that the multiplexing gain of the sum-rate is 2M. In section \ref{sec:cog_int} we explore the effect of cross-over information on the multiplexing gain: we define and demonstrate that the multiplexing gain of the cognitive interference channel (where no cross-over messages are present) is 1, in contrast to the multiplexing gain of 2 seen in the cognitive $X$ channel.  These results allow us to compare the achievable rate regions of the cognitive and cognitive $X$ channels in section \ref{sec:comparison} at various SNRs. We conclude in section \ref{sec:conclusion}.  Due to space limitations, most of the proofs are deferred to \cite{devroye_2x2} which may be found online.
\section{The MIMO cognitive X channel}
\label{sec:cog_X}
% Need to introduce the single antenna, 2 antenna thing 
In this section we show that the MIMO cognitive $X$ channel with $M$ antennas at each of the 4 nodes has a multiplexing gain of $2M$, that of the broadcast channel of Fig. \ref{fig:3Xchannels}(3). %Repeating for clarity, in the cognitive channel, shown in Fig. \ref{fig:channels25} (a),  there are two messages, one from $({\cal S}_1\rightarrow {\cal R}_1)$, and the other from $({\cal S}_2\rightarrow {\cal R}_2)$.  There is no cross-over information from $({\cal S}_1\rightarrow{\cal R}_2)$ or $({\cal S}_2\rightarrow{\cal R}_1)$.
%Here ${\cal S}_2$ knows the message $X_1$, as seen by the directed double arrow in Fig. \ref{fig:channels25}(s).  The multiplexing gain of this channel is 1,the same as the interference channel.
%Consider now the same 2 sender, 2 receiver  Gaussian noise channel as Fig. \ref{fig:channels}(b) except that  here we \emph{do} have cross-over information. That is,  each sender has an independent message destined to each receiver, for a total of four messages,  as shown in the cognitive $X$ channel of Fig. \ref{fig:channels25}(b).  %This is also referred to as the X-channel \cite{jafar_degrees, MMK}.
%${\cal S}_1$ wishes to send message $s_{11} \in \{1,2,\cdots 2^{nR_{11}}\}$, encoded as $M_{11}\in {\cal M}_{11}$  to ${\cal R}_1$ (at rate $R_{11}$) and $s_{12}\in \{1,2,\cdots, 2^{nR_{12}}\}$, encoded as $M_{12}\in {\cal M}_{12}$  to ${\cal R}_2$ (at rate $R_{12}$) in $n$ channel uses.
%Similarly, ${\cal S}_2$ wishes to send message $s_{21} \in \{1,2,\cdots 2^{nR_{21}}\}$, encoded as $M_{21}\in {\cal M}_{21}$ to ${\cal R}_1$ (at rate $R_{21}$) and $s_{22}\in \{1,2,\cdots, 2^{nR_{22}}\}$ encoded as $M_{22}\in {\cal M}_{22}$ to ${\cal R}_2$ (at rate $R_{22}$) in $n$ channel uses.  
The MIMO cognitive X  channel is shown in Fig.  \ref{fig:channels25}(a). The word cognitive stems from the work \cite{devroye_IEEE} in which an interference channel with asymmetric side-information between Txs is called a \emph{cognitive radio channel} (also known as an interference channel with degraded message sets in \cite{jovicic, wu:ifcdms}).  In our figures, we denote side-information using a \emph{double} arrow accompanied by the message that is known a priori. In the MIMO $X$-channel there are four messages: 11, 12, 21 and 22, which are encoded as $M_{11}, M_{12}, M_{21}$ and $M_{22}$ respectively. Our
channel is a standard interference channel \cite{carleial} with direct channel coefficients of 1 and cross-over coefficients of $\alpha_{12}$ and $\alpha_{21}$. The $M$-dimensional transmitted random variables $X_1\in {\cal X}_1$ and $X_2\in {\cal X}_2$  are received as the signals $Y_1\in {\cal Y}_1$ and $Y_2\in {\cal Y}_2$ in the sets according to the conditional distributions $p(y_1|x_1,x_2)$ and $p(y_2|x_1,x_2)$.  We consider an additive white Gaussian noise channel,
\begin{align}
Y_1 &= X_1+\alpha_{21} X_2+N_1  \label{eq:y1} \\
 Y_2 &= \alpha_{12} X_1 + X_2 + N_2, \label{eq:y2} \end{align}
 where $N_1 \sim {\cal N}(0,N_1)$ and $N_2 \sim {\cal N}(0,N_2)$ are independent and we assume individual  average transmit power constraints of $P_1$ ($P_2$) on $X_1$ ($X_2$ resp.).  The directed double arrow from $X_1$ to $X_2$ denotes \emph{partial asymmetric side-information}:  the encoding $M_{11}$ is known to Tx 2. 
 %Notice also that only one of Tx 1's messages is known to Tx 2, that is, only \emph{partial} knowledge is used in the following. We could alternatively have allowed $M_{12}$ to be known at the second transmitter. This would lead to analogous results when indices are permuted. 
% The channel is an additive Gaussian noise channel with independent noise at the receivers, so the received signals are:
%\begin{eqnarray}
%Y_1 &=M_{11}+M_{12}+\alpha_{21}(M_{21}+M_{22})+N_1 \\
%Y_2 &= \alpha_{12}(M_{11}+M_{12})+(M_{21}+M_{22})+N_2.
%\end{eqnarray}

Standard definitions of achievable rates and regions are employed
\cite{cover, devroye_IEEE} .
Although our achievable rate region will be defined for finite alphabet sets, in order to determine an achievable region for the Gaussian noise channel, specific forms of the random variables
described in Thm. \ref{thm:X}  are assumed. As in \cite{costa, gallagher, H+K}, Thm. \ref{thm:X}  can readily be extended to memoryless channels with discrete time and continuous alphabets by finely quantizing the input, output, and interference variables (Gaussian in this case).

We now outline an achievable rate region for the Gaussian MIMO cognitive $X$-channel, which will be used to demonstrate a sum-rate scaling law of $2M$.
 The capacity region of the Gaussian MIMO broadcast channel \cite{weingarten} is achieved using Costa's dirty-paper coding techniques \cite{costa}.  In the MIMO cognitive $X$-channel, at Tx 1, the encodings $M_{11}$ and $M_{12}$  may be jointly generated, for example using a dirty-paper like coding scheme. That is, one message may treat the other as non-causally known interference and code so as to mitigate it. At Tx 2, not only may the encodings $M_{21}$ and $M_{22}$ be jointly designed, but they may additionally use $M_{11}$ as \emph{a-priori} known interference. Thus, Tx 2 could encode $M_{22}$ so as to potentially mitigate the interference $Y_2$ will experience from $M_{11}$ as well as $M_{21}$.
 %We demonstrate an achievable region for the discrete, finite alphabet case in Theorem \ref{thm:X} and look at the achieved rate scalings in the Gaussian noise case, assuming specific forms for all involved variables in Theorem \ref{thm:X}.  
 Let $R_{11}$ be the rate from $M_{11}\rightarrow Y_1$, $R_{12}$ from $M_{12}\rightarrow Y_2$, $R_{21}$ from $M_{21}\rightarrow Y_1$ and $R_{22}$ from $M_{22}\rightarrow Y_2$.
\begin{theorem}
Let $Z$ $\eqdef $ $(Y_1,$ $Y_2,$ $X_1,$ $X_2,$ $M_{11},$ $M_{12},$ $M_{21},$ $M_{22})$, and let ${\cal P}$ be the set of distributions on $Z$ that
can be decomposed into the form
\begin{equation}\begin{array}{c}p(m_{11}|m_{12})p(m_{12})p(m_{21})p(m_{22}|m_{11},m_{21})\\ p(x_1|m_{11},m_{12})p(x_2|m_{11},m_{21},m_{22})\\
p(y_1|x_1,x_2)p(y_2|x_1,x_2),\end{array}\label{eq:density}\end{equation}
where we additionally require $p(m_{12},m_{22})=p(m_{12})p(m_{22})$. For any $Z\in {\cal P}$, let $S(Z)$ be the set of all
tuples $(R_{11}, R_{12}, R_{21}, R_{22})$ of non-negative real numbers such that:
%\begin{small}
%\[ \left. \begin{array}{rl}
%R_{11} & \leq I(M_{11};Y_1|M_{21}) - I(M_{11};M_{12}) \\
%R_{21} & \leq I(M_{21};Y_1|M_{11})\\
%R_{11}+R_{21}&\leq I(M_{11},M_{21};Y_1)-I(M_{11};M_{12}) \end{array} \right\}  \begin{array}{c}\mbox{ MAC }\\(M_{11},M_{21})\\ \searrow \swarrow \\ Y_1\end{array} \]
%\[ \left. \begin{array}{rl}
%R_{12} &\leq I(M_{12};Y_2|M_{22}) \\
%R_{22} &\leq I(M_{22};Y_2|M_{12}) - I(M_{22};M_{11},M_{21})\\
%R_{12}+R_{22}&\leq I(M_{12},M_{22};Y_2)-I(M_{22};M_{11},M_{21})\\
%\end{array}\right\}\begin{array}{c} \mbox{ MAC } \\(M_{12},M_{22}) \\ \searrow \swarrow \\Y_2\end{array}\]
%\end{small}

\begin{small}
\[  \begin{array}{rl}
R_{11} & \leq I(M_{11};Y_1|M_{21}) - I(M_{11};M_{12}) \\
R_{21} & \leq I(M_{21};Y_1|M_{11})\\
R_{11}+R_{21}&\leq I(M_{11},M_{21};Y_1)-I(M_{11};M_{12}) \end{array} \]
%\right\}  \begin{array}{c}\mbox{ MAC }\\(M_{11},M_{21})\\ \searrow \swarrow \\ Y_1\end{array} \]
\[ \begin{array}{rl}
R_{12} &\leq I(M_{12};Y_2|M_{22}) \\
R_{22} &\leq I(M_{22};Y_2|M_{12}) - I(M_{22};M_{11},M_{21})\\
R_{12}+R_{22}&\leq I(M_{12},M_{22};Y_2)-I(M_{22};M_{11},M_{21})\\
\end{array}\]
%\right\}\begin{array}{c} \mbox{ MAC } \\(M_{12},M_{22}) \\ \searrow \swarrow \\Y_2\end{array}\]
\end{small}
Any element in the closure of $\cup_{Z\in {\cal P}}S(Z)$ is achievable. 
\label{thm:X}
\end{theorem}
\begin{proof}
The codebook generation, encoding, decoding schemes and formal probability of error analysis are deferred to the manuscript in preparation \cite{devroye_2x2} which may be found online.
Heuristically, notice that the channel from $(M_{11}, M_{21}) \rightarrow Y_1$ is a multiple access which employs dirty paper coding \cite{costa}, reducing the rate $R_{11}$ by $I(M_{11};M_{12})$ (like in Gel'fand-Pinsker \cite{gelfand} coding).
%However, by \eqref{eq:density} we see that $M_{11}$ and $M_{21}$ are in fact independent, and thus the regular MAC equations hold. 
%$M_{11}$ does use a binning scheme with respect to  $M_{12}$, but this does not alter the $(M_{11},M_{21})\rightarrow Y_1$ MAC equations other than 
Similarly, for the MAC $(M_{12},M_{22})\rightarrow Y_2$ the encodings $M_{12}$ and $M_{22}$ are independent (this is true in particular in the Gaussian case of interest in the next subsection, and so we simplify our theorem by ensuring the condition $p(m_{12},m_{22})=p(m_{12})p(m_{22})$) so that the regular MAC equations hold. The rate $R_{22}$ is subject to a penalty of $I(M_{22};M_{11},M_{21})$  in order to guarantee finding an $n$-sequence $m_{22}$ in the desired bin that is jointly typical with any \emph{given} $(m_{11}, m_{21})$. 
\end{proof}
\subsection{MIMO Cognitive $X$-channel multiplexing gain is $2M$}
%The multiplexing gain of the single antenna Gaussian cognitive  channel was shown to be 1. We now proceed to examine the multiplexing gain of the  cognitive Gaussian X-channel.
%We wish to see how the achievable rate tuple varies as a function of the transmit powers, or equivalently, of the SNRs when the N_1ian noise variance is held fixed. 
%We first set  the number of antennas at all nodes equal to $M=1$, in order to gain intuition, and simplify our result:  
We use the above achievable rate region to show that the sum-rate of the MIMO $X$-channel with  \emph{partial asymmetric side-information} has a multiplexing gain $2M$. 
\begin{figure}
\centerline{\epsfig{figure=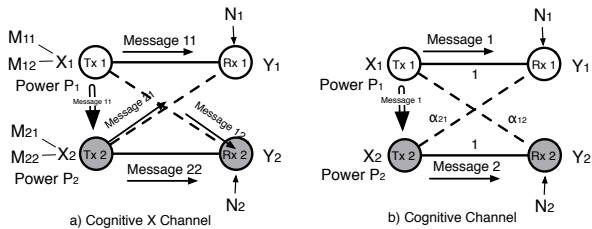, width=8cm}}
\caption{\small \baselineskip=0pt Additive Gaussian noise interference channels with cross-over parameters $\alpha_{12}, \alpha_{21}$, transmitted encodings $X_1, X_2$ with expected transmit power limitations $P_1$ and $P_2$, and  received signals $Y_1$ and $ Y_2$. (a) Cognitive X channel:  four messages encoded as $M_{11}$, $M_{12}$, $M_{21}$, $M_{22}$.  $M_{11}$ is the partial and asymmetric message knowledge at $X_2$. (b) Cognitive channel:two messages encoded as $M_{11}, M_{22}$. $X_1$ is the asymmetric side-information known at $X_2$. }
\label{fig:channels25}
\end{figure}

\begin{corollary}
Consider the MIMO additive Gaussian $X$-channel with partial asymmetric side-information described in eqns. \eqref{eq:y1}, \eqref{eq:y2} and Fig. \ref{fig:channels25}(a) with $P_1$ $= $ $P_2$ $=$ $P$. Then 

\begin{small}
\begin{equation}
\underset{P\rightarrow \infty}{\lim}\frac{\max R_{11}+R_{12}+R_{21}+R_{22}}{\log{P}}=2M,
\end{equation}\end{small}
where the $\max$ is taken over all $(R_{11},R_{12}, R_{21},R_{22})\in
{{\cal C}_{X-cog}}$, where ${\cal C}_{X-cog}$ is the capacity region of the MIMO cognitive $X$-channel.
\end{corollary}

\begin{proof}
We sketch the proof for $M=1$, and defer details, as well as the more involved proof for general $M$ to \cite{devroye_2x2}. Roughly speaking, the general $M$ case is proven by evaluating the same mutual information terms of Thm. \ref{thm:X}, as done for  $M=1$, with the added complications (such as matrix inversions) that arise from considering vectors rather than scalars. %That is, $X_1,X_2, Y_1, Y_2$ are all $M\times 1$ vectors of random variables,  as are $M_{11}, M_{12}, M_{21}, M_{22}, N_1$ and $N_2$.  In order to demonstrate  a multiplexing gain of $2M$, one assumes these random variables are jointly Gaussian, with covariance matrices of specific forms. Their  dependence will be of the same form as the scalar case, where powers will be replaced by covariance matrices, and divisions by matrix inversions, roughly speaking.  Careful matrix manipulation, followed by assuming the covariance matrices to be diagonal (in order to ease the evaluation of determinants) yields the  result. 

First, note that the multiplexing gain of the MIMO broadcast channel, whose capacity region outer bounds ours, with 2 transmit antennas and single receive antennas at the Rxs is 2. 
%Since this channel's capacity region  provides an upper bound to our channel's capacity region, the multiplexing gain cannot exceed 2. 
We will in fact prove that 2 is achievable using the scheme of Thm.  \ref{thm:X}. To do so, we specify forms for the variables, and then optimize the dirty paper coding parameters, similar to Costa's technique \cite{costa}. The Gaussian distributions we assume on all variables are of the form

\begin{small}
\[\begin{array}{ll}
M_{11}  = U_{11}+ \gamma_{1} U_{12}  & U_{11}\sim {\cal N}(0,P_{11})\\
M_{12} = U_{12}, &U_{12} \sim {\cal N}(0,P_{12}) \\
M_{21}=U_{21}, &U_{21}\sim {\cal N}(0,P_{21})\\
M_{22}=U_{22}+\gamma_{2}(U_{21}+a_{12}U_{11}) & U_{22}\sim{\cal N}(0,P_{22}) \\
\end{array}\]
\[\begin{array}{l}
X_1 = U_{11}+U_{12}  \label{eq:dist}\\
X_2 = U_{21}+U_{22}+\sqrt{\frac{(1-\beta)P_2}{P_{11}}}U_{11}\\
Y_1 =\left(1+a_{21}\sqrt{\frac{(1-\beta)P_2}{P_{11}}} \right)\; U_{11}+U_{12}+a_{21}(U_{21}+U_{22})+N_1 \\
Y_2 = \left(a_{12}+\sqrt{\frac{(1-\beta)P_2}{P_{11}}}\right)\;U_{11}+a_{12}U_{12}+(U_{21}+U_{22})+N_2.\end{array}\]
\end{small}
where $P_1 = P_{11}+P_{12}$ and $ \beta P_2 = P_{21}+P_{22}.$
\begin{figure*}
\begin{small}
\begin{align}
%R_{11}&\leq \frac{1}{2}\log_2\left( \frac{P_{11}(P_{11}(1+\theta)^2+P_{12}+\alpha_{21}^2P_{22}+N_1)}{\gamma_1^2(P_{12}(P_{11}(1+\theta)^2+\alpha_{21}^2P_{22}+N_1))-2\gamma_1 P_{11}P_{12}(1+\theta)+P_{11}(P_{12}+\alpha_{21}^2P_{22}+N_1)}\right) \label{eq:R11}\\
%R_{21} &\leq \frac{1}{2}\log_2\left(\frac{P_{21}(P_{11}+\gamma_1^2P_{12}))(P_{11}(1+\theta)^2+P_{12}+a_{21}^2(P_{21}+P_{22})+N_1)}{\gamma_1^2P_{12}(P_{11}(1+\theta)^2+\alpha_{21}^2(P_{21}+P_{22})+N_1)-2\gamma_1P_{11}P_{12}(1+\theta) + P_{11}(P_{12}+\alpha_{21}^2(P_{21}+P_{22})+N_1)}\right) \label{eq:R21}\\
R_{11}+R_{21}& \leq \frac{1}{2}\log_2\left( \frac{P_{11}(P_{11}(1+\theta)^2+P_{12}+\alpha_{21}^2(P_{21}+P_{22})+N_1)}{\gamma_1^2(P_{12}(P_{11}+\alpha_{21}^2P_{22}+N_1))-2\gamma_1 P_{11}P_{12}+P_{11}(P_{12}+\alpha_{21}^2P_{22}+N_1)}\right)\label{eq:R1}
\end{align} 
\end{small}
\end{figure*}
\begin{figure*}
\begin{small}
\begin{align}
%R_{12}&\leq \frac{1}{2}\log_2\left( \frac{\gamma_2^2(a_{12}^2P_{12}+P_{22}+N_2)(P_{21}+(a_{!2}+\theta)^2P_{11})+\gamma_2(-2P_{22}(P_{21}+\eta^2P_{11}))+P_{22}(P_{11}\eta^2+a_{12}^2P_{12}+P_{22}+N_2)}{\gamma_2^2(P_{11}\eta^2+P_{21})(P_{22}+N_2)+\gamma_2(-2P_{22}(P_{11}\eta^2+P_{21})+P_{22}(P_{11}\eta^2+P_{21}+N_2)}\right)\label{eq:R12}\\
%R_{22} &\leq \frac{1}{2}\log_2\left(\frac{(P_{11}\eta^2+P_{21}+P_{22}+N_2)(\gamma_1^2P_{12}(P_{22}+\gamma_2^2\eta^2P_{11})+P_{11}P_{22})}{(P_{11}+\gamma_1^2P_{12}))\gamma_2^2(P_{11}\eta^2+P_{21})(P_{22}+N_2)+\gamma_2(-2P_{22}(P_{11}\eta^2+P_{21})+P_{22}(P_{11}\eta^2+P_{21}+N_2)}\right)\label{eq:R22}\\
R_{12}+R_{22} &\leq \frac{1}{2}\log_2\left(\frac{(P_{11}\eta^2+a_{12}^2P_{12}+P_{21}+P_{22}+N_2)(\gamma_1^2P_{12}(P_{22}+\gamma_2^2\eta^2P_{11})+P_{11}P_{22})}{(P_{11}+\gamma_1^2P_{12}))\gamma_2^2(P_{11}\eta^2+P_{21})(P_{22}+N_2)+\gamma_2(-2P_{22}(P_{11}\eta^2+P_{21})+P_{22}(P_{11}\eta^2+P_{21}+N_2)}\right)\label{eq:R2}
\end{align} 
\end{small}
\end{figure*}

Here the variables $U_{11}, U_{12}, U_{21}, U_{22}$ are all independent, encoding the four messages to be transmitted.  The $\beta$ parameter divides power at Tx 2:  $\beta P_2$ is used in transmitting its own messages, $m_{21}$ and $m_{22}$, while the remainder of the power, $(1-\beta)P_2$ is used to reinforce the message encoded as $m_{11}$ of Tx 1. Notice that because we only assume partial message knowledge (only \emph{one} of the two messages of Tx 1 is known at Tx 2), only one of these messages gets reinforced at Tx 2.  Also note that $p(m_{11},m_{21})=p(m_{11})p(m_{21})$ and $p(m_{12},m_{22})=p(m_{12})p(m_{22})$ as needed in Thm. \ref{thm:X}.  The rates $R_1=R_{11}+R_{21}$ and $R_2=R_{12}+R_{22}$ to each Rx can be calculated separately. Each can be maximized with respect to the relevant dirty-paper coding parameter ($\gamma_{1}$ for Tx 1, and $\gamma_{2}$ for Tx 2). The bounds of Thm. \ref{thm:X} may be evaluated by combining the appropriate determinants of sub-matrices of the overall covariance matrix $E[\Theta \Theta^T]$ where $\Theta \eqdef (M_{11},M_{21},M_{12},M_{22},Y_1,Y_2)$. The six bounds of Thm. \ref{thm:X} are evaluated explicitly  in \cite{devroye_2x2} online; we simply demonstrate the resulting expressions for the sum-rate to Rx 1, $R_1$ as well as to Rx 2, $R_2$ in the expressions \eqref{eq:R1} and \eqref{eq:R2} resp.
%Evaluation of the bounds under this choice of Gaussian random variables yields eqns. \eqref{eq:R11}-\eqref{eq:R1} and \eqref{eq:R12}-\eqref{eq:R2}. There, 
In order to simplify the expressions, we have set
\[{\small  \theta \eqdef a_{21}\sqrt{\frac{(1-\beta)P_2}{P_{11}}} \;\;\; \mbox{ and } \;\;\; \eta \eqdef (a_{12}+\theta).}\]
We could search for the $\gamma_1$ and $\gamma_2$ that jointly optimize $R_1+R_2$. However, noticing that $R_1$ depends only on $\gamma_1$, we heuristically select  $\gamma_1$ to maximize $R_1$, as 
\begin{equation} {\small \gamma_1 = \frac{P_{11}(1+\theta)}{P_{11}(1+\theta)^2+\alpha_{21}^2P_{22}+N_1}.\label{eq:gamma1}}\end{equation}
When we substitute this $\gamma_1$ into the bounds on Rx 1's total sum-rate  we obtain the bound \eqref{eq:R1sub}.
%\begin{figure*}
%\begin{align}
%R_{11}&\leq \frac{1}{2}\log_2\left( 1+\frac{P_{11}}{\alpha_{21}^2 P_{22}+N_1}\right) \\
%R_{21} &\leq \frac{1}{2}\log_2\left(1+\frac{\alpha_{21}^2P_{21}(P_{11}P_{12}+(P_{11}+\alpha_{21}^2P_{22}+N_1)^2)}{(\alpha_{21}^2P_{22}+N_1)(P_{12}+P_{11}+\alpha_{21}^2P_{22}+N_1)(P_{11}+\alpha_{21}^2P_{22}+N_1)}\right) \\
\begin{figure*}
\begin{equation} R_{11}+R_{21}\leq \frac{1}{2}\log_2\left( \frac{(P_{11}(1+\theta)^2+P_{12}+\alpha_{21}^2(P_{21}+P_{22})+N_1)(P_{11}(1+\theta)^2+\alpha_{21}^2P_{22}+N_1)}{(\alpha_{21}^2P_{22}+N_1)(P_{11}(1+\theta)^2+P_{12}+\alpha_{21}^2P_{22}+N_1)}\right)\label{eq:R1sub} \end{equation} \end{figure*}
%\end{align} 
%\end{figure*}
Notice that an important cancellation occurs in the denominator of \eqref{eq:R1sub} when the optimal $\gamma_1$ is substituted in the sum-rate expression.%The bounds for the second receiver work out to

To maximize the  $R_1$ bound, we had set $ \gamma_1$ as in \eqref{eq:gamma1}. 
Although we could maximize $R_2$ with respect to $\gamma_2$, we use a simpler and more heuristic approach and simply minimize the denominator of the sum-rate $R_2=R_{12}+R_{22}$ with respect to $\gamma_2$, which yields $\gamma_2 = \frac{P_{22}}{P_{22}+N_2}.$
It is interesting to note that this is exactly the same dirty paper coding parameter as Costa derives. It is intuitively pleasing, and although it does not strictly maximize $R_2$ with respect to $\gamma_2$, as we will see shortly, it performs sufficiently well in the limit of large SNR, thus performs asymptotically optimally. If we fix  $P_{22}$ (does not scale with $P$) and let $P_{11} = P_{12} = P_{21}$ all scale like $P$, subject to $P_{11}+P_{12} = P$ and $P_{21}+P_{22}=P$, then the bound on the total sum rate to both Rxs $R_{11}+R_{12}+R_{21}+R_{22}$  scales like $2\log P$. This can be seen by noting that $\gamma_2$ remains fixed and $\gamma_1\rightarrow 1$ as $P\rightarrow \infty$.  
Keeping $P_{22}$ fixed was crucial for achieving the $\log P$ scaling in $R_1$. Intuitively, this is because of the asymmetric message knowledge; the interference Tx 2 causes the Tx 1 is not mitigated. Keeping $P_{22}$ constant still allows Tx 2 to dirty paper code, or mitigate the interference caused by $M_{11}$ and $M_{21}$ to Rx 2's signal $Y_2$, while causing asymptotically (as $P_{11}, P_{12}, P_{21}\rightarrow \infty$) negligible interference to $Y_1$.  We note that the sum-rate scaling does not depend on the power division parameter $\beta$. That is, the sum-rate scaling remains 2 for any $\beta > 0$ (meaning some power is in fact allocated to transmit the messages of Tx 2, and this power will then $\rightarrow \infty$). 
%We will illustrate that the sum-rate scaling result holds when $\beta=1$, that is,  when all Tx 2's power is allocated towards its own messages and none towards amplifying Tx 1's messages.  
%Notice also that  we did not optimize $\gamma_1$ and $\gamma_2$ jointly. Rather, we selected $\gamma_1$ so as to maximize $R_{11}+R_{21}$ and then selected $\gamma_2$ so as to minimize the \emph{denominator only} of $R_{12}+R_{22}$. 
The joint optimization of $\gamma_1, \gamma_2$ could lead to higher rates in lower SNR regimes. However, at high SNR our choice of parameters leads to an optimal sum-rate scaling. 
\end{proof}
\section{The cognitive interference channel}
\label{sec:cog_int}
\begin{figure*}
\begin{tabular}{ccc}
\epsfig{figure=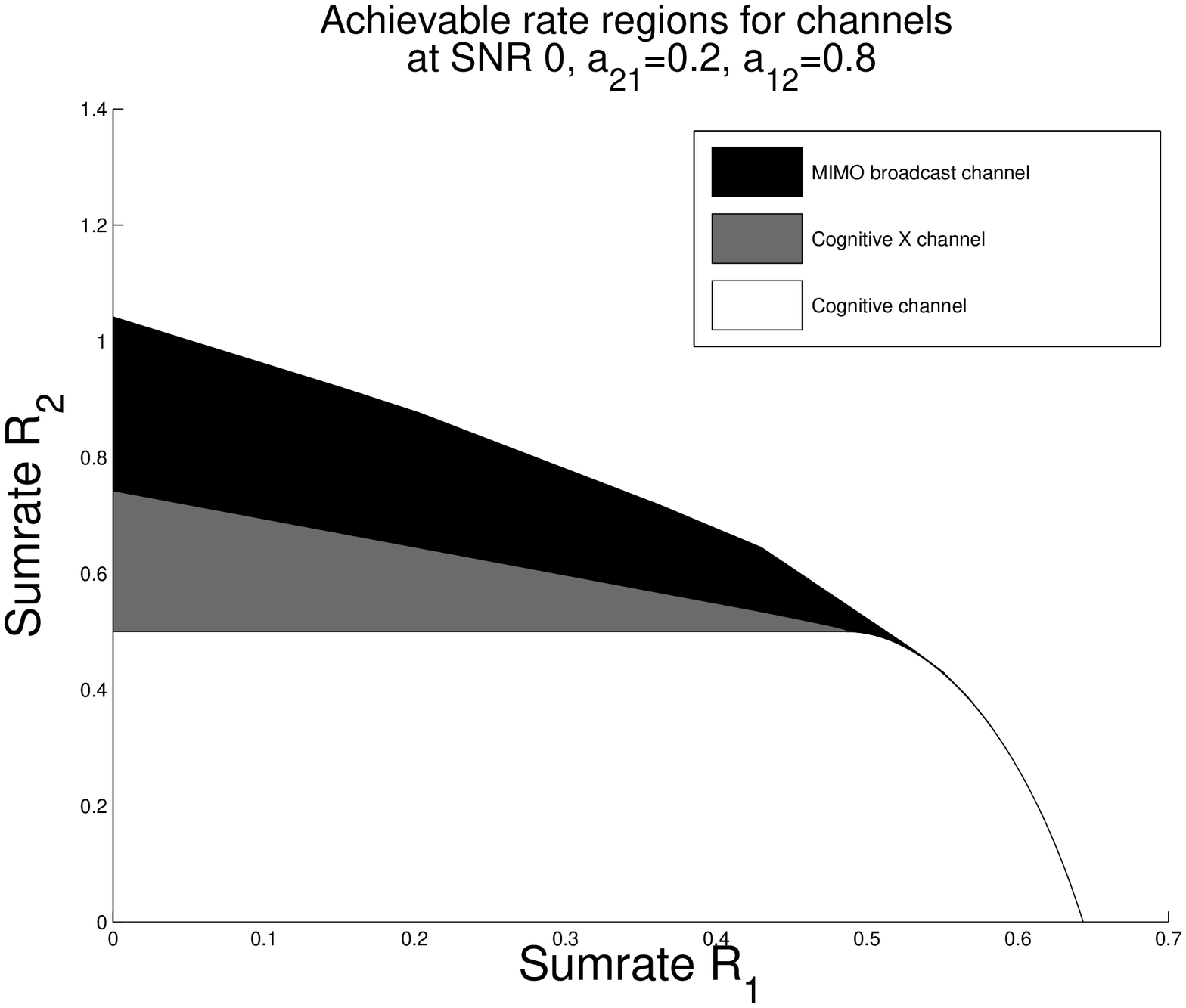 , width=5cm}&
\epsfig{figure=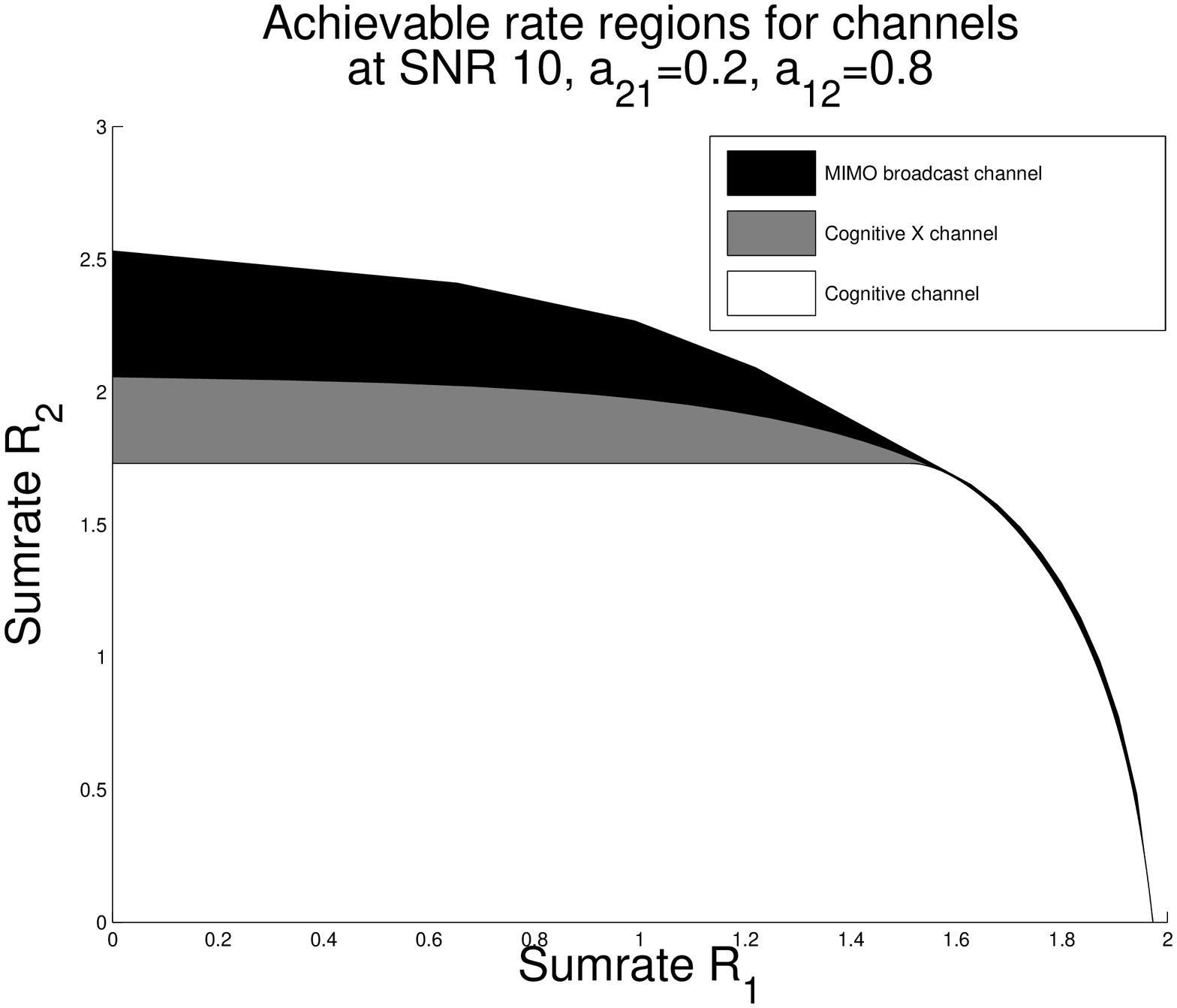 , width=5cm} &
\epsfig{figure=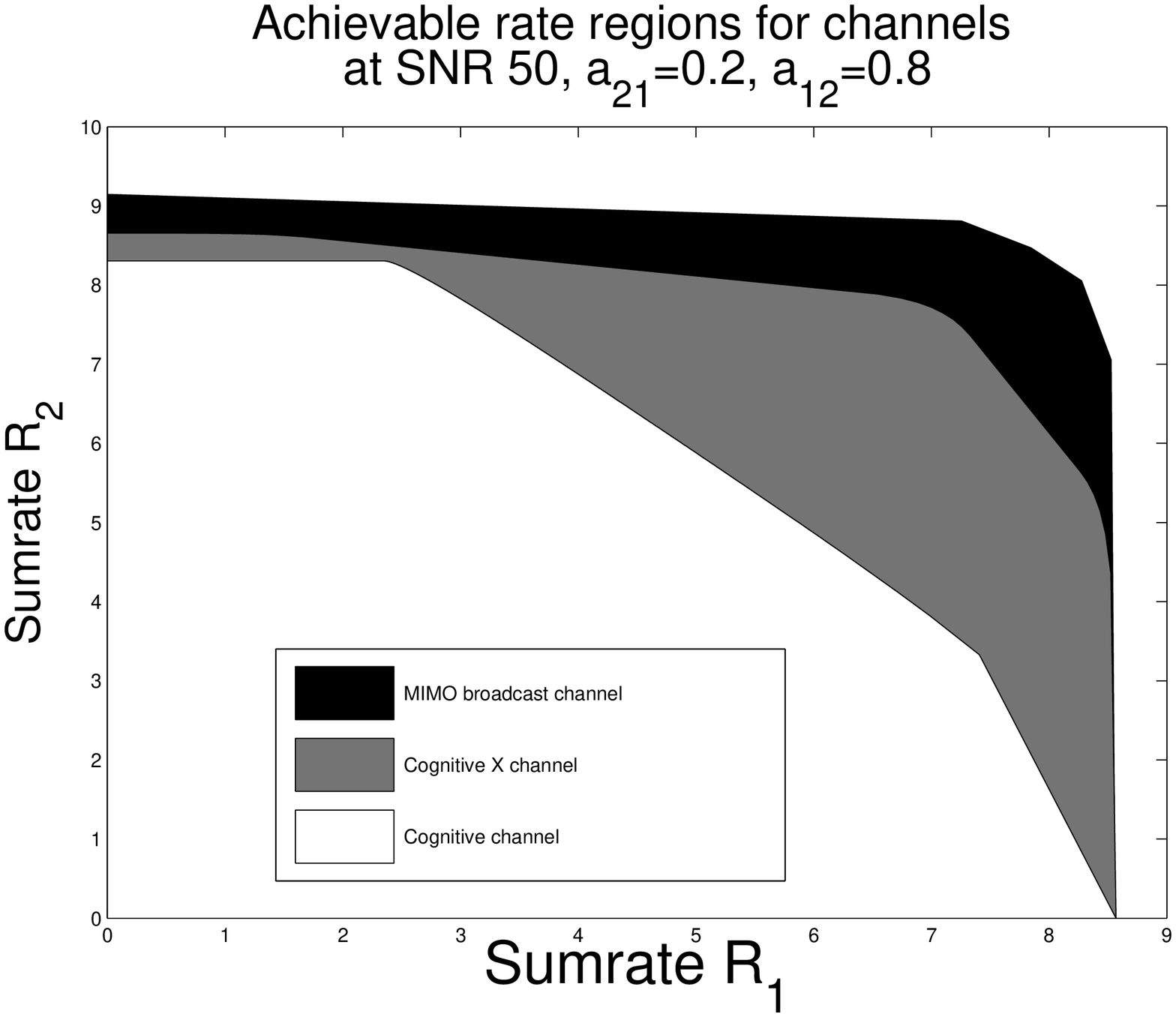 , width=5cm} \\
(a) & (b)&(c)\\
SNR = 0dB & SNR=10dB &  SNR=50dB
\end{tabular}
\caption{Comparison of the cognitive interference and the cognitive $X$ channels at various SNRs.}
\label{fig:SNRs}
\end{figure*}
% FIX: needs to be compared to something, so briefly review the interference and cognitive results
In the previous section we demonstrated that the scaling law of the sum-rate of the MIMO cognitive $X$ channel, with \emph{partial transmitter side-information}  is $2M$, which is optimal in the limit as SNR $\rightarrow \infty$.  In this section we investigate whether partial asymmetric side-information is always equivalent to full symmetric transmitter side information in terms of sum-rate scaling as SNR $\rightarrow \infty$. To do so we look at another channel with partial asymmetric side information at the transmitters: the recently explored cognitive interference channel (also known as the interference channel with degraded message sets \cite{jovicic} or the cognitive radio channel \cite{devroye_IEEE}), shown in Fig. \ref{fig:channels25}(b). We consider the same additive Gaussian noise channel as in \eqref{eq:y1}, \eqref{eq:y2}.  The only difference with the cognitive $X$-channel is the absence of cross-over messages 12 and 21.  
We will see that while partial asymmetric side information in the $X$-channel results in the same sum-rate scaling as a fully cooperative (at the transmitters) $X$-channel, the opposite is true of partial asymmetric side information in the interference channel: at high SNR its sum-rate scales like the interference channel. In other words, although partial side-information may help
the interference channel in a medium ${\rm SNR}$-regime \cite{devroye_IEEE, jovicic}, at high ${\rm SNR}$, one cannot improve the scaling law of the sum-rate. 
% This extended the results of \cite{jafar_degrees} in which the scaling law of the MIMO $X$-channel, which uses no Tx side-information, was shown to lie in the range $\left[\lfloor \frac{4M}{3} \rfloor,\frac{4M}{3}\right]$. 
%In this section we investigate the effect of cross-over information in channels with partial transmitter side-information. Specifically, we concentrate on the multiplexing gains of the cognitive interference \cite{devroye_IEEE, jovicic} and the cognitive $X$ channels. 
%This channel, show has 2 messages: 11 and 22, while the latter has 4 messages: the same two direct messages 11 and 22, as well as the cross-over messages 12 and 21. Interestingly, partial side-information improves the sum-rate scaling of the $X$ channel from $4/3$ (at best) to $2$, while partial side-information in the interference channel achieves the same unit sum-rate scaling law as the interference channel without side-information. 
%We consider the same additive Gaussian noise channel as in \eqref{eq:y1}, \eqref{eq:y2}. 
%We show that the sum-rate of the Gaussian interference channel with partial side information (also the cognitive radio channel, or the interference channel with degraded message sets), shown in Fig. \ref{fig:channels25}(a) with two independent messages Tx 1 $\rightarrow$ Rx 1 and Tx 2$\rightarrow$Rx 2,  scales at best like $\log P$ (not $2 \log P$). In other words, although partial side-information may help
%the interference channel in a medium ${\rm SNR}$-regime \cite{devroye_IEEE, jovicic}, at high ${\rm SNR}$, one cannot improve the scaling law of the sum-rate.  
The Gaussian cognitive interference  channel considered here is the same channel as that of \cite{jovicic}, where its capacity region is derived for the case $\alpha_{21}\leq 1$, and sum-rate capacity is found for $\alpha_{21}>1$. The next theorem is a direct result from this capacity region. 
\begin{theorem}
Consider the Gaussian interference channel %$\alpha_{21}\leq 1$,
 where Tx 2 has non-causal knowledge of the message of Tx 1, described in eqns. \eqref{eq:y1}, \eqref{eq:y2} and Fig. \ref{fig:channels25}(b) with $P_1=P_2=P$. Then %the sum-rate capacity of this channel satisfies
\begin{equation}
\underset{P\rightarrow \infty}{\lim}\frac{\max_{(R_1,R_2)\in
{\cal{C}}}{R_1+R_2}}{\log{P}}=1,
\end{equation}
where $R_i$ corresponds to the rates from the $i$-th
source to the $i$-th Rx, and $\cal{C}$ is the capacity
region of the channel.
\thmend \label{thm:logP}
\end{theorem}

The proof of this result is deferred to work \cite{devroye_2x2} which may be viewed online. It employs eqns. (24), (25) and Corollary 4.1 of \cite{jovicic}. 
This theorem shows that the cognitive channel of Fig. \ref{fig:channels25}(b) behaves like the interference channel at high SNR. Thus, when partial side-information is available at the Txs, although achievable rates may be enhanced in the low and medium SNR range as compared to the interference channel, at large SNR, the two channels scale in the same manner. 
This is in sharp contrast to the cognitive $X$ channel, where partial asymmetric side information results in the same sum-rate scaling as  the fully cooperative broadcast channel at high SNR. 
\section{Comparison of Cognitive and Cognitive $X$ channel regions at various SNRs}
\label{sec:comparison}
In this section, we numerically evaluate the capacity region of the cognitive channel of Fig. \ref{fig:channels25}(b) and \cite{jovicic}  and compare it with the achievable region of the cognitive $X$ channel described in Thm.  \ref{thm:X} and Fig. \ref{fig:channels25}(a) under our choice of variables, as well as the MIMO broadcast channel with 2 Tx antennas and 2 single antenna receivers. In doing so, we highlight the dependence of the rate region, and in particular the sum-rate scaling, on the SNR. At 0 and 10dB receive SNR, the two regions almost coincide for large $R_1$. At high SNR (50dB) the sum-rate scaling increases, and the gap between the sum-rate achieved by the cognitive interference and the cognitive $X$-channels widens, confirming the sum-rate scaling laws of 1 and 2 respectively. Fig. \ref{fig:SNRs} contrasts the achievable rate regions for cross-over parameters $\alpha_{12}=0.8, \;\alpha_{21}=0.2$ at the SNRs 0 and 50 dB. Notice that the broadcast channel always forms an outer bound, but that the achievable rate region appears to be tight as for large $R_1$. We speculate the achievable rate region is tight  for areas of the rate region where the slope of the tangent is smaller than -1).
\section{Conclusion}
\label{sec:conclusion}
In this paper we have shown that the multiplexing gain of the sum-rate of the MIMO  cognitive $X$ channel is $2M$, achieving the optimal sum-rate scaling. Particularly surprising, and pleasing, is that the optimal sum-rate scaling may be achieved with only partial asymmetric side-information at the Txs. This could lead one to think that  partial asymmetric side-information, rather than full side-information (or cooperation) between Txs always yields the optimal sum-rate scaling. However, we then showed that this is not the case: in the interference channel with asymmetric side-information, where no cross-over information is permitted, the scaling of the sum-rate for the single antenna channel is 1, rather than the optimal 2. This demonstrates that at large SNR,  asymmetric cooperation does not give any gain, in terms of scaling, over an interference channel in which there is no Tx cooperation. 

\baselineskip=2pt
\bibliographystyle{IEEEtranS}
\bibliography{mybib}

\end{document}